\documentclass[a4paper,11pt]{article}

\usepackage[a4paper, margin=2cm]{geometry}
\usepackage[utf8]{inputenc}
\usepackage[T1]{fontenc}
\usepackage{cite}
\usepackage{amsmath}
\usepackage{amsfonts}
\usepackage{amssymb}
\usepackage{amsthm}
\usepackage{amsbsy}
\usepackage{bm}
\usepackage{mathrsfs}
\usepackage{slashed}
\usepackage{graphicx}
\usepackage{tikz-cd}
\usepackage{fancyhdr}
\usepackage{jheppub}

\newcommand{\dd}{\mathrm{d}}

\newcommand{\mw}{\mathrm{w}}

\newcommand{\pp}{\mathrm{p}}

\newcommand{\da}{\dot{\alpha}}
\newcommand{\jj}{\mathrm{j}}

\newcommand{\pd}{\partial}
\newcommand{\al}{\alpha}
\newcommand{\bt}{\beta}
\newcommand{\ad}{\dot{\alpha} }
\newcommand{\bd}{\dot{\beta}}
\newcommand{\bra}{\langle}
\newcommand{\ket}{\rangle}
\newcommand{\bx}{\square}
\newcommand{\tc}{\tilde{c}}
\newcommand{\tb}{\tilde{b}}

\title{\boldmath On the Spectrum and Spacetime Supersymmetry of Heterotic Ambitwistor String}

 \author{ Henrique Flores }
 \author{and Matheus Lize }
 \affiliation{\textit{ICTP South American Institute for Fundamental Research,\\
Instituto de F\'{i}sica Te\'{o}rica, Universidade Estadual Paulista, \\
Rua Dr. Bento Teobaldo Ferraz 271, S\~{a}o Paulo, SP Brasil}}

\emailAdd{henrique.flores@unesp.br}
\emailAdd{ matheus.lize@unesp.br }

\abstract{We analyse the BRST cohomology of the Ramond sector of heterotic
ambitwistor theory. We also write the free string field theory action and
compute the supersymmetry transformations.}

\begin{document}
\maketitle
\flushbottom

\section{Introduction.}

The ambitwistor string was introduced by Mason and Skinner in \cite{Mason:2013sva} as a string theory whose tree amplitudes
reproduce the Cachazo-He-Yuan formulas for massless scattering \cite{Cachazo:2014}.
It contains only left-moving worldsheet fields and has no massive states.
As described in \cite{Mason:2013sva}, it is the $\alpha' \to 0$ limit of superstring theory,
even though only the $GSO(+)$ sector of the Type II version correctly describes $d=10$ $N=2$ supergravity.

The spectra was identified in \cite{Berkovits:2018jvm}
using the standard BRST method where equations of motion and gauge invariances are
derived from the cohomology at ghost-number 2 of the BRST operator.
The quadratic term for the string field theory action was also constructed in a gauge-invariant manner;
and except for the $GSO(+)$ sector of the Type II ambitwistor string, 
the kinetic terms for the other ambitwistor strings are non-unitary, \textit{i.e.}  they contain more than two derivatives. 
In this paper, we focus on the Ramond sector of heterotic ambitwistor string.
We want to understand if the fermionic spectra of the theory is given by
non-unitary equations of motion,
and study the supersymmetry transformations of these non-unitary systems.

Section \ref{desnecessaria} reviews the ambitwistor model and sets our notation.
It can be skipped on a first reading.

We start in section \ref{cohomology}, where
we use the standard BRST method to compute the equations of motion of the Ramond sector for the heterotic system.
These represent the fermionic degrees of freedom of the theory,
and our analysis shows that they also follow non-unitary equations of motion.
We write a gauge-invariant version of theory in terms of Fronsdal fields\cite{Fang:1978wz}.
The kinetic term of the fermionic ambitwistor string field theory action is also computed in section \ref{saction}.
It is expressed in terms of gauge-invariant objects and resembles Fronsdal's free action despite having more
derivatives.

Finally, in section \ref{susy} we write the supersymmetry transformations of the system.
In RNS language, the supersymmetry operator is defined on-shell and thus gives
the supersymmetry transformations up to equations of motion.
Then we prove the invariance of the action under supersymmetry
transformations.

\section{Ambitwistor Action and Ramond Sector.} \label{desnecessaria}

We first review the ambitwistor model.
Its main purpose is to set the basic definitions and notation.

The heterotic ambitwistor model is defined by the free action

\begin{equation}
S = \frac{1}{2\pi} \int \dd^{2}z
\bigg( \pp_{m} \bar{\partial} x^{m}
+ \psi_{m} \bar{\partial} \psi^{m}
+ b \bar{\partial} c + \tb \bar{\partial} \tc
+ \beta \bar{\partial} \gamma + S_{\jj} \bigg)
\label{ahet}
\end{equation}

\noindent
where $\pp_{m}$ is a worldsheet holomorphic one-form
and $x^{m}$ is an holomorphic coordinate function.
The $b$ and $c$ fields
together with $\beta$ and $\gamma$ are the Faddeev-Poppov ghosts of
superconformal worldsheet symmetry.
Particular to the heterotic model, we have the current action $S_{\jj}$;
its specific form is irrelevant for us,
we only require the existence of a current $\jj^{a}$
with conformal weight $1$ that satisfies the OPE

\begin{equation}
\jj^{a}(z) \jj^{b}(w) \sim \frac{\delta^{ab}}{(z-w)^{2}}
+ \frac{f^{ab}_{c} \jj^{c}(w)}{(z-w)} ,
\end{equation}

\noindent
being $f^{ab}_c$ the structure constants of the Lie algebra in question.
The Ambitwistor model differs from the superstring
due to the presence of the $\tb$ and $\tc$ ghosts
related to the gauge symmetries of the light-cone constrain: $\pp^{2} = 0$.
These ghosts have conformal weights $2$ and $-1$ respectively and
both are worldsheet fermions.

Our Majorana spinors $\psi^{m}$ will be rewritten in the complex linear
combinations:

\begin{align} \label{bosopsi}
\psi^{\pm i} = \frac{1}{\sqrt{2}} \left( \psi^{2i - 1} \mp \psi^{2i} \right)
\end{align}

\noindent
for $i=1,\ldots,5$ that are subsequently bosonized to

\begin{equation} \label{pbosonization}
\psi^{\pm i} (z) = \exp \bigg( \pm \phi_{i} (z) \bigg) c_{\pm e_{i}}
\end{equation}

\noindent
with $\phi$'s satisfying

\begin{equation}
\phi_{i} (z) \phi_{j} (\mw) \sim
+ \, \delta_{ij} \ln (z - \mw)
\end{equation}

\noindent
The $(\beta, \gamma)$ system is bosonized with extra fermions $(\xi, \eta)$\cite{Friedan:1985ge},
both primaries of conformal weight $0$ and $1$ respectively:

\begin{equation}
\beta = \partial \xi e^{- \phi_{6}} c_{e_{6}} \quad \ \text{and} \quad \
\gamma = \eta e^{\phi_{6}} c_{e_{6}}.
\end{equation}

\noindent
This choice follows the conventions of \cite{Kostelecky:1986xg} and \cite{Koh:1987hm} where
we have introduced the cocycles $c_{e_{i}}$ and $c_{e_{6}}$.
During the computation of cohomology, cocycle factors are important and must be 
taken into account.
The definition of cocycles depends on the way
we order the different $\phi_i$.
For us the chiral bosons corresponding to $\psi^{m}$ are
ordered from $1$ to $5$ while the boson coming from the $\beta \gamma$ system
is labeled as $6$.
A review of how to operate with cocycles can be found in 
\cite{Kostelecky:1986xg} and a brief explanation is written in appendix \ref{appA}.
The sixth boson has OPE:

\begin{equation}
\phi_{6} (z) \phi_{6} (\mw) \, \sim \, - \ln (z - \mw)
\end{equation}

\noindent
while $(\xi, \eta)$ form a free system:

\begin{equation}
\xi(z) \eta(\mw) \sim \frac{1}{(z-\mw)} 
\end{equation}

The symmetries of this action are encoded in the following
BRST charge:

\begin{equation} \label{bosonizedQ}
Q = \oint \frac{dz}{2\pi i} \Bigg[ c \bigg(  \mathrm{T}_{\text{matter}} + T_{\tb\tc} + T_{\beta\gamma} + \mathrm{T}_{\jj} \bigg)
+ bc\pd c + \frac{1}{2}\tilde{c} \pp^{2} + \gamma \pp^{m} \psi_{m}  - \gamma^{2}\tilde{b} \Bigg]
\end{equation}

\noindent
provided

\begin{subequations}

\begin{equation}
\mathrm{T}_{\text{matter}} =
- \pp_{m} \partial x^{m}
+ \frac{1}{2}\sum_{i=1}^{5}  \pd \phi_{i}\pd \phi_{i},
\quad \quad
\mathrm{T}_{\tilde{b}\tilde{c}}
= \tc \partial \tb -2 \tb \partial \tc,
\end{equation}

\begin{equation}
T_{\beta\gamma} = -\frac{1}{2}\pd\phi_{6}\pd\phi_{6} - \pd^{2}\phi_{6} - \eta\pd \xi,
\quad \texttt{and} \quad   
\gamma^{2} = \eta\pd \eta e^{+2\phi_{6}}.
\end{equation}
\end{subequations}

\noindent
These are all the stress-energy tensors for $(x^{m}, \pp_{m}, \psi^{m})$,
$(\beta, \gamma)$ and $(\tb, \tc)$.
We only require for the stress tensor of the current sector,
$\mathrm{T}_{\jj}$,
that the following OPE is satisfied:

\begin{equation}
\mathrm{T}_{\jj}(z) \mathrm{T}_{\jj}(w) \sim
\frac{c_{j}}{2(z-w)^{4}}
+ \frac{2 \mathrm{T}_{\jj}(w)}{(z-w)^{2}}
+ \frac{\partial \mathrm{T}_{\jj}(w)}{(z-w)}.
\end{equation}

\noindent
Then, provided the central charge of the current system is $16$,
it is possible to show
that $Q^{2} = 0$ when the spacetime is
$10$-dimensional.


\section{Cohomology.} \label{cohomology}

In this section, we compute the ghost number $2$ BRST cohomology of
the Ambitwistor string for states in the Ramond sector.
The cohomology of the Neveu-Schwartz sector has already been computed in \cite{Berkovits:2018jvm}.

We start by writing the most general vertex operator
and the most general gauge parameter.
Once all equations of motion and gauge transformations are obtained, we solve
the algebraic gauge conditions to obtain a set of independent field equations.

\subsection{Vertex operators.}

States are defined by picture number $-1/2$ and ghost number $2$
BRST cohomology. We define ghost and picture numbers by the expressions:

\begin{equation}
N_{\text{ghost}} = -\oint \frac{\dd z}{2 \pi i} \bigg( bc +\tb\tc + \xi \eta \bigg)
\label{18}
\quad \text{and} \quad
N_{\text{picture}} = \oint \frac{\dd z}{2\pi i} \, \bigg( \xi \eta- \pd \phi_{6} \bigg).
\end{equation}

\paragraph{Vertex Operator.}
The most general ghost number $2$ and picture number $-1/2$ vertex operator that is
annihilated by $b_{0}$ is given by the sum,

\begin{equation} \label{vertex}
V_{R} = V_{+} + V_{-},
\end{equation}

\noindent
where $V_{+}$ and $V_{-}$ are the $GSO(+)$ and $GSO(-)$
combinations. 
The $GSO(+)$ vertex operator is given by:

\begin{align} \label{vertexop}
V_{+} &=
c  \eta  S^{\alpha} e^{\phi/2} \mathbf{A}_{\alpha} +
\tc \eta  S^{\alpha} e^{\phi/2} \mathbf{B}_{\alpha} +
c \tc {S}^{\da} e^{-\phi/2} \partial x^{m} {\mathbf{C}}_{m\da} +
c \tc {S}^{\da}e^{-\phi/2} \mathrm{p}_{m} {\mathbf{D}}^{m}_{\da}+\\
&
+c \tc {S}^{\da} e^{-\phi/2} \jj^{a} {\mathbf{E}}^{a}_{\da}
\nonumber 
+ c \partial \tc {S}^{\da} e^{-\phi/2} {\mathbf{F}}_{\da}
+ c \tc {S}^{\da}\partial e^{-\phi/2} {\mathbf{G}}_{\da}+\\
&
+ c \tc \psi_{m}(\slashed{\psi} S)^{\alpha} e^{-\phi/2} \mathbf{H}^{m}_{\alpha}
+ c \tc \partial \tc \partial \xi S^{\alpha} e^{-3\phi/2} \mathbf{I}_{\alpha}
+ \tc \partial \tc {S}^{\da} e^{-\phi/2} {\mathbf{J}}_{\da}
\end{align}


\noindent
while $V_{-}$ is obtained from $V_{+}$ by changing the chirality of our spinors.
Notice that the vertices $\psi^{m} \psi^{n} {S}^{\ad}$ and $\partial {S}^{\ad}$
have not been written. In bosonized form, these combinations are related to $\psi \slashed{\psi} {S}$
via field redefinitions\cite{Kostelecky:1986xg};
there is no need to worry about them.

\paragraph{Gauge vertex.} As for the gauge transformations, we parametrize them
by ghost number $1$ and picture number $-1/2$
vertex operators:

\begin{equation} \label{gvertex}
\Lambda = c  {S}^{\ad}e^{-\phi_{6}/2} {\lambda}_{\ad}
+ \tc   {S}^{\ad}e^{-\phi_{6}/2}{ \omega}_{\ad}
+ c \tc  \partial \xi S^{\alpha}e^{-3\phi_{6}/2} \mu_{\alpha}.
\end{equation}

\noindent
Both expressions \eqref{vertex} and \eqref{gvertex}
constitute the basic field content of BRST cohomology.

\subsection{Equations of motion and gauge symmetries.}

For clarity we consider only the $GSO(+)$ sector.
The $GSO(-)$ is obtained by replacing chiral indices for anti-chiral and vice-versa.
We present the equations of motion organized by ghost number as they were obtained from the OPE of $Q$ and
$V_{+}$. We also write the worldsheet operator that multiplies the resulting equation of motion.

\begin{itemize}
\item For  $ ( 2 c , 1 \tilde{c}) $ multiplying $( {S}^{\ad}e^{-\phi_{6}/2}c \tc \pd^{2}c )$:
\end{itemize}

\begin{equation} \label{eomfirst}
+ \frac{1}{2}\pd_{m}{\mathbf{D}}_{\ad }^{m} +{\mathbf{F}}_{\ad } -\frac{3}{8}{\mathbf{G}}_{\ad } - \frac{9}{4}(\Gamma^{m})_{\ad}^{\bt}\mathbf{H}_{m\bt} = 0
\end{equation}

\begin{itemize}
\item For $( 0 c , 1 \tilde{c} )$ multiplying $( {S}^{\ad}e^{3\phi_{6}/2}\tc \eta \pd \eta)$:
\end{itemize}

\begin{equation}
+ {\mathbf{J}}_{\ad } - \frac{i}{\sqrt{2}}(\Gamma^{m})_{\ad}^{\bt}\pd_{m}\mathbf{B}_{\bt } = 0
\end{equation}

\begin{itemize}
\item For $( 1 c , 0 \tilde{c} ) $ multiplying $(  {S}^{\ad}e^{3\phi_{6}/2}c\eta \pd \eta  )$:
\end{itemize}

\begin{equation}
-\frac{i}{\sqrt{2}}(\Gamma^{m})_{\ad}^{\bt}\pd_{m} \mathbf{A}_{\bt}  -{\mathbf{G}}_{\ad } + {\mathbf{F}}_{\ad } = 0
\end{equation}

\begin{itemize}
\item For $( 1 c , 2 \tilde{c} )$

\begin{itemize}
\item multiplying $(  S^{\alpha}e^{-\phi_{6}/2}c \tc \pd \tc \pp_{m} )$:
\end{itemize}

\begin{equation}
-\frac{1}{2}\bx {\mathbf{D}}_{\ad }^{m} +{\mathbf{C}}_{\ad}^{m}
- \pd^{m}{\mathbf{F}}_{\ad }
- \frac{i}{\sqrt{2}}(\Gamma^{m})_{\ad}^{\bt}\mathbf{I}_{\bt} = 0
\end{equation}

\begin{itemize}
\item multiplying $(  S^{\ad}e^{-\phi_{6}/2}c \tc \pd^{2} \tc  )$:
\end{itemize}

\begin{equation}
-\frac{1}{2}\pd^{m} {\mathbf{C}}_{m\ad} - {\mathbf{J}}_{\ad} = 0
\end{equation}

\begin{itemize}
\item multiplying $(  {S}^{\ad}e^{-\phi_{6}/2}c \tc \pd \tc \pd x^{m} )$:
\end{itemize}

\begin{equation}
-\frac{1}{2}\bx{\mathbf{C}}_{m\ad} -\pd_{m}{\mathbf{J}}_{\ad } = 0
\end{equation}

\begin{itemize}
\item multiplying $(   {S}^{\ad}e^{-\phi_{6}/2}c \tc \pd \tc \pd\phi_{6} )$:
\end{itemize}

\begin{equation}
+\frac{1}{4}\bx {\mathbf{G}}_{\ad }
+ \frac{1}{2}{\mathbf{J}}_{\ad }
+ \frac{i}{\sqrt{2}}(\Gamma^{m})_{\ad}^{\bt}\pd_{m}\mathbf{I}_{\bt} = 0
\end{equation}

\begin{itemize}
\item multiplying $\left(  c \tc \pd \tc \psi^{m}(\slashed{\psi} {S} )^{\al} \right)$:
\end{itemize}

\begin{equation}
+ \frac{1}{2}\bx \mathbf{H}_{\al}^{m}
- \frac{i}{4\sqrt{2}}\pd_{m}\mathbf{I}_{\al}
+ \frac{i}{8\times 9\sqrt{2}} (\Gamma_{m})^{\bd}_{\al}(\slashed{\pd}\mathbf{I})_{\bd}
+ \frac{1}{9\times 4}(\Gamma_{m})^{\bd}_{\al}{\mathbf{J}}_{\bd} = 0
\end{equation}
\end{itemize}

\begin{itemize}
\item For $ 1 c , 1 \tilde{c} $

\begin{itemize}
\item multiplying $(  S^{\alpha}e^{\phi_{6}/2}c\eta \pd \tc  )$:
\end{itemize}

\begin{equation}
- \frac{1}{2}\bx \mathbf{A}_{\al} + \mathbf{B}_{\al}
+ 2 \mathbf{I}_{\alpha}
- \frac{i}{\sqrt{2}}(\Gamma^{m})_{\al}^{\bd}\pd_{m}{\mathbf{F}}_{\bd}  = 0
\end{equation}

\begin{itemize}
\item multiplying $(  S^{\alpha}e^{\phi_{6}/2}c\tc\eta \pd x^{m} )$:
\end{itemize}

\begin{equation}
- \pd_{m}\mathbf{B}_{\al }
+ \frac{i}{\sqrt{2}}(\Gamma^{n})_{\al}^{\bd}\pd_{n}{\mathbf{C}}_{m\bd}  = 0
\end{equation}

\begin{itemize}
\item multiplying $(  S^{\alpha}e^{\phi_{6}/2}c\tc \eta \pp_{m} )$:
\end{itemize}

\begin{equation}
- \pd^{m}\mathbf{A}_{\al}
+ \frac{i}{\sqrt{2}}(\Gamma^{n})_{\al}^{\bd}\pd_{n}{\mathbf{D}}_{\bd}^{m}
+ \frac{i}{2\sqrt{2}}(\Gamma^{m})_{\al}^{\bd}{\mathbf{G}}_{\bd }
- \frac{i8}{\sqrt{2}}\mathbf{H}^{m}_{\al }
- \frac{i}{\sqrt{2}} \mathbf{H}_{\bt n} (\Gamma^{n})^{\bt}_{\ad} (\Gamma^{m})^{\ad}_{\al} = 0
\end{equation}

\begin{itemize}
\item multiplying $( S^{\alpha}e^{\phi_{6}/2}c\tc \pd \eta  )$:
\end{itemize}

\begin{align}
- \mathbf{B}_{\al }
+ 3 \mathbf{I}_{\alpha}
+ \frac{i}{\sqrt{2}}(\Gamma^{m})_{\al}^{\bd}{\mathbf{C}}_{m\bd}
- \frac{i}{2\sqrt{2}}(\Gamma^{m})_{\al}^{\bd}\pd_{m}{\mathbf{G}}_{\bd }
+ \frac{8i}{\sqrt{2}}\pd^{m}\mathbf{H}_{\al m} +\\\nonumber
+ \frac{i}{\sqrt{2}} (\Gamma^{n})^{\bd}_{\al} (\Gamma^{m})^{\tau}_{\bd}\pd_{n}\mathbf{H}_{\tau m} = 0
\end{align}

\begin{itemize}
\item multiplying $( S^{\alpha} e^{\phi_{6}/2}c\tc \eta\pd \phi_{6} )$:
\end{itemize}

\begin{align}
\frac{1}{2} \mathbf{B}_{\al }
+ 4 \mathbf{I}_{\alpha}
+ \frac{i}{\sqrt{2}}(\Gamma^{m})_{\al}^{\bd}{\mathbf{C}}_{m\bd}
- \frac{i}{\sqrt{2}}(\Gamma^{m})_{\al}^{\bd}\pd_{m}{\mathbf{G}}_{\bd }
+ \frac{8i}{\sqrt{2}}\pd^{m}\mathbf{H}_{\al m}+\\\nonumber
+ \frac{i}{\sqrt{2}} (\Gamma^{n})^{\bd}_{\al} (\Gamma^{m})^{\tau}_{\bd}\pd_{n}\mathbf{H}_{\tau m} = 0
\end{align}

\begin{itemize}
\item multiplying $\left( \eta c\tc \psi^{m}(\slashed{\psi}S)^{\ad}e^{\phi_{6}/2} \right)$:
\end{itemize}

\begin{align} \label{eomlast}
+ \frac{i}{2\sqrt{2}} \Bigg[ \frac{1}{4} \pd_{m}{\mathbf{G}}_{\bd }
- \frac{1}{9\times 8} (\Gamma_{m})^{\tau}_{\ad}(\slashed{\pd}{\mathbf{G}})_{\tau} \Bigg]
- \frac{i}{\sqrt{2}} \Bigg[\frac{1}{4} {\mathbf{C}}_{\ad m}
- \frac{1}{9\times 8} (\Gamma_{m})^{\tau}_{\ad}(\mathbb{\slashed{{C}}})_{\tau} \Bigg]+
\nonumber \\
\nonumber \\
- \frac{i}{\sqrt{2}} \Bigg[ -\frac{1}{4}(\Gamma^{n})^{\bt}_{\ad}\pd_{m}\mathbf{H}_{\bt n}
- (\Gamma^{n})^{\bt}_{\ad}\pd_{n}\mathbf{H}_{\bt m} + \frac{1}{9}(\Gamma_{m})^{\bt}_{\ad}\pd^{n}\mathbf{H}_{\bt n} \Bigg] +\\ \nonumber
+ \frac{1}{36} (\Gamma_{m})^{\al}_{\ad} \mathbf{B}_{\al }- \frac{i}{\sqrt{2}} \Bigg[ \frac{1}{9\times8}(\Gamma_{m})^{\bt}_{\ad}(\Gamma^{l})^{\bd}_{\bt}(\Gamma^{p})^{\tau}_{\bd}\pd_{l}\mathbf{H}_{\tau p} \Bigg] =0
\end{align}
\end{itemize}

\noindent
These 14 equations of motion are all invariant under the following $10$ gauge transformations:

\begin{subequations} \label{gt}
\begin{align} 
&\delta \mathbf{A}_{\alpha} =+ \frac{i}{\sqrt{2}}(\Gamma^{m})^{\bd}_{\al}\pd_{m}{\lambda}_{\bd}
+2 \mu_{\alpha}
\\
&\delta \mathbf{B}_{\alpha} =  +\frac{i}{\sqrt{2}}(\Gamma^{m})^{\bd}_{\al}\pd_{m}{\omega}_{\bd}
\\
&\delta \mathbf{I}_{\alpha}=\frac{1}{2}\bx \mu_{\alpha}
\\
&\delta\mathbf{H}^{m}_{\al} = \frac{1}{9\times4}(\Gamma^{m})_{\al}^{\bd}{\omega}_{\bd} + \frac{i}{4\sqrt{2}}\pd_{m}\mu_{\al} - \frac{i}{8\times 9\sqrt{2}} (\Gamma_{m})^{\bd}_{\al}(\slashed{\pd}\mu)_{\bd}
\\
&\delta {\mathbf{C}}_{m\ad} = \pd_{m}{\omega}_{\ad}
\\
&\delta {\mathbf{D}}_{\ad}^{m}=\pd_{m}{\lambda}_{\ad}-\frac{i}{\sqrt{2}} \, (\Gamma^{m})_{\ad}^{\beta}  \mu_{\beta}
\\
&\delta {\mathbf{E}}_{\ad}^{A}=0
\\
&\delta {\mathbf{F}}_{\ad}=-\frac{1}{2}\bx{\lambda}_{\ad} + {\omega}_{\ad}
\\
&\delta {\mathbf{G}}_{\ad}={\omega}_{\ad}-\frac{2i}{\sqrt{2}} \, (\Gamma^{m})_{\ad}^{\beta}\pd_{m}\mu_{\beta}
\\
&\delta {\mathbf{J}}_{\ad}=-\frac{1}{2}\bx {\omega}_{\ad}
\end{align}
\end{subequations}

We determined the basic content of ghost number $2$ BRST cohomology;
all equations of motion have been written between \eqref{eomfirst}
and \eqref{eomlast}. This set is highly redundant, and the next
step is to use \eqref{gt} to stablish the independent field equations. 

\subsection{Gauge-fixing and independent equations of motion.}

In order to find the independent set of equations of motion, we
begin by fixing algebraic gauge conditions and solving auxiliary field equations.
Let us gauge-fix $\mathbf{A}$ and ${\mathbf{F}}$ to zero
using the parameters $\mu$ and $\omega$,
that is, we choose $\mu = -\mathbf{A}$ and $\omega = - \mathbf{{F}}$ so
that the residual gauge parameters $\mu'$ and $\omega'$ must
satisfy:

\begin{equation} \label{gaugecondition1}
\mu_{\alpha}^{'} + \frac{i}{2\sqrt{2}}(\Gamma^{m})^{\bd}_{\al}\pd_{m}{\lambda}_{\bd} = 0,
\end{equation}

\noindent
and

\begin{equation} \label{gaugecondition2}
{\omega}_{\ad}^{'} - \frac{1}{2}\bx{\lambda}_{\ad} = 0.
\end{equation}

\noindent
After this gauge fixing, the following auxiliary field conditions can be imposed:

\begin{subequations}
\begin{align}
& {\mathbf{G}}^{m}_{\ad} = 0,
\\
\nonumber \\
&\mathbf{B}_{\al} = -2 \mathbf{I}_{\al},
\\
\nonumber \\
& {\mathbf{C}}^{m}_{\ad} = + \frac{1}{2}\bx {\mathbf{D}}^{m}_{\ad}  +\frac{i}{\sqrt{2}} \, (\Gamma^{m}\mathbf{I})_{\ad},
\\
\nonumber \\
& {\mathbf{J}}_{\ad} =
-\frac{1}{4} \bx \, \pd_{m} {\mathbf{D}}^{m}_{\ad}  -\frac{i}{2\sqrt{2}}\, (\slashed{\pd}\mathbf{I})_{\ad},
\\
\nonumber \\
& \mathbf{H}^{m}_{\al} =  \frac{1}{8} \slashed{\pd}^{\bd}_{\al}{\mathbf{D}}^{m}_{\bd}
- \frac{1}{18\times 8} (\Gamma^{m}\Gamma_{n} \, \slashed{\pd})^{\bd}_{\al} {\mathbf{D}}_{\bd}^{n}.
\end{align}
\end{subequations}

\noindent
At this point it is already clear that there only remains two independent fields
given by $\mathbf{D}^{m}_{\ad}$ and $\mathbf{I}_{\al}$.
Moreover, the only remaining gauge parameter is ${\lambda}$.
We leave the gluino field ${\mathbf{E}}^{a}_{\bd}$ out of the discussion
since its equation of motion is already the Dirac equation and it has no gauge transformations.

Finally, the following set of $3$ equations,

\begin{subequations}
\begin{equation}
\frac{i}{\sqrt{2}} \partial^{m} \mathbf{I}_{\al} =
\square \Bigg( \frac{1}{4} \slashed{\pd}^{\bd}_{\al} \mathbf{D}^{m}_{\bd}
- \frac{1}{12} (\Gamma^{m})^{\bd}_{\al} \partial_{n} \mathbf{D}^{n}_{\bd} \Bigg)
\end{equation}

\begin{equation}
2 \partial_{m} {\mathbf{D}}^{m}_{\ad}
+ (\Gamma_{n} \Gamma_{p})^{\bd}_{\ad} \partial^{n} {\mathbf{D}}^{p}_{\bd}  = 0
\end{equation}

\begin{equation}
\slashed{\pd}^{\ad}_{\al}{\mathbf{E}}_{\ad}^{a} =0
\end{equation}
\end{subequations}

\noindent
with the corresponding gauge transformations:

\begin{subequations}
\begin{equation}
\delta {\mathbf{D}}^{m}_{\ad}= \frac{3}{4}\pd^{m}{\lambda}_{\ad}
- \frac{1}{4} (\Gamma^{mn})^{\;\;\bd}_{\ad}\pd_{n}{\lambda}_{\bd}
\end{equation}

\begin{equation}
\delta \mathbf{I}_{\al} =- \frac{i}{4\sqrt{2}}\slashed{\pd}^{\bd}_{\al}\bx{\lambda}_{\bd}\\
\end{equation}
\end{subequations}

\noindent
defines the spectrum of the theory.

\paragraph{Gauge-invariant description.}
Consider the following field redefinitions:

\begin{subequations}\label{redefine}
\begin{equation}
{\mathbf{d}}^{m}_{\ad} = {\mathbf{D}}^{m}_{\ad} 
- \frac{1}{6} (\Gamma^{m})^{\al}_{\ad} \slashed{\mathbf{D}}_{\al}
\end{equation}

\begin{equation}
\mathbf{i}_{\al} = +\frac{i4}{\sqrt{2}} \mathbf{I}_{\al}  
+\frac{1}{6}\bx{\slashed{\mathbf{D}}}_{\al}
\end{equation}
\end{subequations}

\noindent
such that our gauge transformations are mapped to

\begin{equation}
\delta {\mathbf{d}}^{m}_{\ad} = \partial^{m} {\lambda}_{\ad}
\quad \texttt{and} \quad
\delta \mathbf{i}_{\al} = 0.
\end{equation}

\noindent
The gauge-invariant object is then naturally defined as:

\begin{equation}
\mathbf{F}_{mn\ad} = \pd_{m}{\mathbf{d}}_{n\ad} - \pd_{n}{\mathbf{d}}_{m\ad}
\end{equation}

\noindent
which allows us to write the equations of motion in the following form:

\begin{subequations} \label{reom}
\begin{equation}
\pd_{m} \mathbf{i}_{\al} = \bx \mathbf{F}_{m\al}
\end{equation}

\noindent
and

\begin{equation}
(\slashed{\mathbf{F}})_{\ad}=0
\end{equation}
\end{subequations}

\noindent
where

\begin{equation}
\mathbf{F}_{m\al} \equiv (\Gamma^{n})^{\ad}_{\al} \mathbf{F}_{mn\ad} 
= (\slashed{\pd} {\mathbf{d}}_{m} 
- \pd_{m} {\slashed{\mathbf{d}}})_{\al}.
\end{equation}

\noindent
In the formulation of free higher-spin theories
$\mathbf{F}_{m}$ is called Fronsdal tensor\cite{Fang:1978wz},
it is the analog of the Ricci curvature in spin $2$ formulation.

This section started with the most general ghost number $2$ picture $-1/2$
vertex operator. Then we obtained all equations of motion from the BRST method 
together with all gauge transformations parametrized by ghost number $1$ picture
$-1/2$ vertex operators.
By fixing some of this gauge freedom, we have found a independent set of
equations of motion that can be parametrized by Fronsdal fields. 
The next natural step is to write the spacetime action that gives the dynamics of this 
system.

\section{Action} \label{saction}

The kinetic term of the ambitwistor string field theory was defined in
\cite{Berkovits:2018jvm}:

\begin{equation} \label{action}
S[V] = \bra I \circ V^{(-3/2)}(0) \, \pd c \, Q  V^{(-1/2)}(0) \ket,
\end{equation}

\noindent
where $V^{-1/2}$ is the vertex operator \eqref{vertex} introduced in the previous section,
an element of the small Hilbert space that is also constrained to satisfy $L_{0} V = b_{0} V = 0 $.
The RNS string has one additional feature: the picture number.
It is necessary to saturate the background charge of supermoduli space to $-2$,
and that is why we need a string field with picture
$-1/2$, $V^{-1/2}$, together with a string field with picture $-3/2$, $V^{-3/2}$.
We define picture raising, $Z$, and picture lowering, $Y$, by the following expressions:

\begin{equation} \label{raising}
Z = c \pd \xi  + e^{\phi_{6}} \pp_{m} \psi^{m}  -\pd (e^{2\phi_{6}} \eta \tb) - e^{2\phi_{6}} \pd\eta \tb ,
\end{equation}

\begin{equation}\label{lowering}
Y(z) = \tc \pd \xi e^{-2\phi_{6}},
\end{equation}

\noindent
so that we can obtain $V^{-3/2}$ from $V^{-1/2}$ via

\begin{equation}
V^{-3/2}(z) =  \frac{1}{2\pi i} \oint \frac{dw}{(w-z)} Y(w) V^{-1/2}(z).
\end{equation}

\noindent
Using the auxiliary gauge-fixing conditions imposed on the previous section,
we obtain

\begin{align}
&V^{-3/2} =\\
&+ \tc \pd \tc  S^{\al}e^{-3\phi_{6}/2} \mathbf{B}_{\alpha}
- c \tc \pd \tc \pd \xi {S}^{\ad}e^{-5\phi_{6}/2} \partial x^{m} {\mathbf{C}}_{m\ad}
- c \tc \pd \tc \pd \xi {S}^{\ad}e^{-5\phi_{6}/2} \mathrm{p}_{m} {\mathbf{D}}^{m}_{\ad}
\nonumber \\
&- c \tc\pd \tc \pd \xi {S}^{\ad}e^{-5\phi_{6}/2} j^{a} {\mathbf{E}}^{a}_{\ad}
- c \tc \pd \tc \pd \xi \psi_{m}(\slashed{\psi} S)^{\al}e^{-5\phi_{6}/2} \mathbf{H}^{m}_{\alpha}
- \frac{1}{2} c \tc \partial \tc \partial \xi \partial^{2} \tc \partial^{2} \xi S^{\al}e^{-7\phi_{6}/2} \mathbf{I}_{\alpha}
\end{align}

\noindent
The composition $I \circ V^{-3/2}$ is the BPZ
conjugate of the picture $-3/2$ field with $I= - 1/z$.
We should be careful when computing the conformal transformation $I \circ V^{-3/2}$
because $V^{-1/2}$ is not primary.
From the OPE with the stress-energy tensor

\begin{equation}
T(z) V^{-1/2}(0) \sim z^{-3} {S}^{\ad}e^{-\phi_{6}/2}c \tc \left(
\frac{1}{2}\pd_{m}{\mathbf{D}}_{\ad }^{m} +{\mathbf{F}}_{\ad } -\frac{3}{8}{\mathbf{G}}_{\ad }
- \frac{9}{4}\slashed{\mathbf{H}}_{\ad}\right) + \, \cdots
\end{equation}

\noindent
we obtain a cubic pole contribution that changes the finite conformal transformation to

\begin{equation}
I \circ V =
\Bigg[ V\big(I(z)\big) +
\frac{1}{2} \frac{ I''(z) }{ [ I'(z) ]^{2}}\;
\# \big(I(z)\big) \Bigg].
\end{equation}

\noindent
where $\#$ is cubic pole coefficient.
Even after the auxiliary conditions are imposed 
we still have non-primary contributions that must be taken into account.

To calculate the free action,
we fix the normalization $\bra c \pd c \pd^{2}c \tc \pd \tc \pd^{2}\tc e^{-2\phi_{6}}\ket = 4$,
then the correlation function \eqref{action} gives the following
gauge-invariant action:

\begin{equation} \label{raction}
S_{R} = -\int d^{\,10}x \left[  \frac{1}{2}\mathbf{d}^{m\al}\bx\left( \mathbf{F}_{m\al}
-\frac{1}{2}(\gamma_{m})_{\al\bt}\slashed{\mathbf{F}}^{\bt}\right)
+\frac{1}{2}(\slashed{\mathbf{F}})^{\al}\mathbf{i}_{\al}
-\frac{i}{2}\text{Tr}\bigg( {\mathbf{E}}\slashed{\pd}{\mathbf{E}} \bigg)
\right].
\end{equation}

\noindent
In this expression  we used the symmetric gamma matrices 
$(\gamma^{m}_{\alpha \beta},\gamma_{m}^{\alpha \beta})$
defined in appendix \ref{appA}. 
When using these symmetric matrices, the charge conjugation is used to 
eliminate all dotted indices; different chiralities are just represented by upper and lower
indices, \textit{i.e.} $( C^{\al \ad} \mathbf{d}^{m}_{\ad} = \mathbf{d}^{m\al})$. 

We have written a non-unitary action that gives the equations of motion obtained in \eqref{reom}.
It closely resembles the gauge-invariant formulation of spin $3/2$, the difference being the presence
of more derivatives. 
Let us proceed and study the supersymmetry of this non-unitary system.

\section{Supersymmetry.} \label{susy}

Let us define the supersymmetry generator as

\begin{equation} \label{susyg}
\mathbf{Q}_{\al}^{-1/2} =\frac{1}{2\pi i} \oint \dd z \, {S}_{\al}e^{-\phi_{6}/2}
\end{equation}

\noindent
Notice that it carries picture, which means that supersymmetry algebra only closes on-shell.
We need the picture $1/2$ supersymmetric charge:

\begin{equation} \label{susyg2}
\mathbf{Q}_{\al}^{1/2} = \frac{1}{2\pi i} \oint \dd z
\bigg[ i \pp_{m} (\gamma^{m})_{\al\bt}S^{\bt}e^{\phi_{6}/2} + \tb \eta {S}_{\al}e^{3\phi_{6}/2} \bigg].
\end{equation}

\noindent
to obtain $\{ Q^{-1/2}_{\al} , Q^{1/2}_{\beta} \} = 2\gamma^{m}_{\al \bt} \pp_{m} $.
In practice, supersymmetry transformations are written up to equations of motion.
One also needs to choose a GSO sector
to have well-defined supersymmetry transformations,
otherwise there will be branch cuts.
Given the generator \eqref{susyg}, we need use the
GSO(+) vertex operator. 

\subsection{Supersymmetry transformations of NS and R sectors.}

The Neveu-Schwarz vertex operator in picture $-1$ was written in
\cite{Berkovits:2018jvm}:

\begin{align} \label{het}
&V_{NS}^{-1} = \nonumber\\
& e^{-\phi_{6}} c\tilde{c} \bigg[ \left( G^{(1)}_{(mn)} + B^{(1)}_{[mn]} \right) \, \pp^{m} \psi^{n} 
+ \left( G^{(2)}_{(mn)} + B^{(2)}_{[mn]} \right) \pd x^{m}\psi^{n}
+ C_{mnp} \psi^{m}\psi^{n}\psi^{p}  
+ \jj^{a} \psi^{m} A^{a}_{m} \bigg]
 \nonumber \\ 
&+ e^{-\phi_{6}} c\tilde{c} \pd \psi^{m} A_{m}^{(4)} 
+ \pd\phi_{6} e^{-\phi_{6}} c\tilde{c} A^{(3)}_{m} \psi^{m}
+ \pd \xi e^{-2\phi_{6}}\pd^{2}\tilde{c}\tilde{c}cS^{(4)}
+ \eta c S^{(1)}
+ \pd \xi e^{-2\phi_{6}} \pd^{2}cc\tilde{c}S^{(2)}
\nonumber \\
&+ \dots 
\end{align}

\noindent
where $\dots$ depends only on the previous fields.
In \cite{Berkovits:2018jvm}, 
the fields $(B^{(1)}_{mn},A^{(3)}_{m},A^{(4)}, S^{(1)},S^{(2)})$ of
\eqref{het} were gauged to zero. If we choose to keep this gauge, we must observe
that in general supersymmetry does not preserve a given gauge condition. 
Therefore when calculating supersymmetry transformations, we have to choose 
the gauge parameter $\Lambda$:
 
\begin{equation} \label{susygauge}
\delta_{\zeta}V_{NS}^{-1} = \bigg[\zeta\mathbf{Q}^{-1/2},V_{R}^{-1/2}\bigg]
+\bigg[ Q_{BRST}, \Lambda^{-1} \bigg],
\end{equation}

\noindent
which is a vertex operator of ghost number $1$ and picture $-1$, to ensure that
$\delta_{\zeta}(B^{(1)}_{mn},A^{(3)}_{m},A^{(4)}$,
$ S^{(1)},S^{(2)})$ all give zero.
In the transformations below, the contributions 
of $\mathbf{H}$ are due to the gauge-fixing of these auxiliary fields:

\begin{align}
&\delta_{\zeta} G^{(1)}_{mn} =2(\zeta\gamma_{(m}{\mathbf{D}}_{n)})
\\
&\delta_{\zeta} G^{(2)}_{mn}=\frac{2}{5}(\zeta\gamma_{(n}{\mathbf{C}}_{m)})
-\frac{48}{5} \pd_{(n}\zeta\mathbf{H}_{m)} 
\\
&\delta_{\zeta} B^{(2)}_{mn}=-4(\zeta\gamma_{[n}{\mathbf{C}}_{m]})
-\frac{48}{5} ( \zeta\pd_{[m}\mathbf{H}_{n]} ) 
\\
&\delta_{\zeta} C_{mnp} = \frac{3}{2}\pd_{[p}(\zeta\gamma_{m}{\mathbf{D}}_{n]})
-24(\zeta\gamma_{[np}\mathbf{H}_{m]})+6(\zeta\gamma_{mnp}\slashed{\mathbf{H}})
\end{align}

\noindent
and using the field redefinitions of \cite{Berkovits:2018jvm}:

\begin{equation} \label{susyns}
h_{mn} = G^{(1)}_{mn} + \frac{1}{4}\eta_{mn}h^{r}_{r},
\quad
t = \frac{1}{4}\bx h^{m}_{m} + G^{m(2)}_{m}
\quad \texttt{and} \quad
B^{(2)}_{mn} = B_{mn}
\end{equation}

\noindent
we arrive at 

\begin{align} \label{susyr}
\delta_{\zeta} h_{mn} &= 2 \zeta \gamma_{(m}{\mathbf{d}}_{n)}\\
\delta_{\zeta} t &=  \zeta \mathbf{i}\\
\delta_{\zeta} C_{mnp} &= - 3(\zeta\gamma_{t[mn}\mathbf{F}^{t}_{\; p]}) 
- 3(\zeta\gamma_{[m}\mathbf{F}_{n p]})\\
\delta_{\zeta} B_{mn} &  = -2\bx (\zeta \gamma_{[m}\mathbf{d}_{n]})
-  (\zeta \gamma_{mn}\mathbf{i})
+ \frac{1}{6} (\zeta \gamma_{mn}\pd_{p} \mathbf{F}^{p})\\
\delta_{\zeta} A_{m}^{a} & = \frac{i}{2}(\zeta \gamma_{m} {\mathbf{E}}^{a}).
\label{NStrans}
\end{align}

\noindent
The term $(\zeta \gamma_{mn}\pd_{p} \mathbf{F}^{p} )$ is zero if we use the equation of motion
$\slashed{\mathbf{F}} =0$, and so could not have been obtained from the supersymmetry generator
\eqref{susyg}. This term was added by hand in order to make the 
action invariant under supersymmetry. 


For the Ramond sector the same can be done if we use instead the
picture $+1/2$ supersymmetry generator \eqref{susyg2}:

\begin{align}
\delta_{\zeta} {\mathbf{d}}_{m}^{\al} &=
+  (\gamma^{rs}\zeta)^{\al}\pd_{s}h_{mr}
-2(\gamma^{np}\zeta)^{\al} C_{mnp} +
\frac{1}{3}(\gamma_{mnps}\zeta) C^{nps}   \\
\delta_{\zeta} \mathbf{i}_{\al} &= 2  (\zeta\slashed{\pd})_{\al} t
- (\gamma^{mnp}\zeta)_{\al} H_{mnp}  +\frac{1}{3} (\gamma^{mnp}\zeta)_{\al} \bx C_{mnp}\\
\delta_{\zeta} {\mathbf{E}}^{a\bt} &= -\frac{1}{4}F_{mn}(\gamma^{mn}\zeta)^{\bt}
\label{Rtrans}
\end{align}

At this point, we have obtain the supersymmetry transformations of both 
NS and R system for the independent fields of the theory in equations \eqref{susyr}
to \eqref{susyns}.
Let us proceed and check that indeed the total $GSO(+)$ action is supersymmetric
invariant.

\subsection{Supersymmetry invariance of the action.}


The action that describes the Neveu-Schwarz sector is

\begin{align} \label{NSaction}
S_{NS} = -\int d^{10} x &\left[ \frac{1}{2} h^{mn}\bx\left( R_{mn}
-\frac{1}{2}\eta_{mn}R\right) - tR +\frac{1}{4}  \text{Tr}(F^{mn}F_{mn}) +\right.
\nonumber \\
&\left. \qquad \qquad -C^{mnp}H_{mnp}
+ \frac{1}{2}C^{mnp}\left(\bx C_{mnp}  - \frac{1}{2}\pd_{[p}\pd^{r}C_{mn]r} \right) \right]
\end{align}

\noindent
where $H_{mnp}$ is the field strength for $B_{mn}$ and $R_{mn}$ is the Ricci tensor.
This expression is equivalent to the action written in equation $(4.13)$ of \cite{Berkovits:2018jvm}
if we shift $t$ by $t \mapsto t + R^{2}$. 
The equations of motion derived from \eqref{NSaction} are

\begin{align}
&\bx R_{mn} - \pd_m\pd_n t =0 \;, \quad R =0 \;, \quad \bx C_{mnp} - H_{mnp} =0\;,
\nonumber \\
\nonumber \\
&\qquad \qquad \pd^{m}C_{mnp}=0 \;, \quad \texttt{and} \quad \partial_{m} F^{mn} = 0.
\end{align}

\noindent
Now, the Ramond sector is described by equation \eqref{raction}:

\begin{equation}
S_{R} = -\int d^{\,10}x \left[  \frac{1}{2}\mathbf{d}^{m\al}\bx\left( \mathbf{F}_{m\al}
-\frac{1}{2}(\gamma_{m})_{\al\bt}\slashed{\mathbf{F}}^{\bt}\right)
+\frac{1}{2}(\slashed{\mathbf{F}})^{\al}\mathbf{i}_{\al}
-\frac{i}{2}\text{Tr}\bigg( {\mathbf{E}}\slashed{\pd}{\mathbf{E}} \bigg)
\right].
\end{equation}

\noindent
from which we obtain the following set of equations of motion -- \eqref{reom}:

\begin{equation}
\pd_{m} \mathbf{i}_{\al} = \bx \mathbf{F}_{m\al}\;,
\quad
\slashed{\mathbf{F}}^{\al}=0 
\quad
\texttt{and}
\quad
i \slashed{\partial}_{\alpha \beta} \mathbf{E}^{a \, \beta}  = 0.
\end{equation}

From now on, 
we leave the Yang-Mills system out of the discussion because its
supersymmetry transformations and action are already standard.
For later use, let us write the supersymmetry transformation for all field strengths: 

\begin{subequations}
\label{susyfs}
\begin{align}
\delta_{\zeta}  R_{mn} &= (\zeta\pd_{(m} \mathbf{F}_{n)})
+ (\zeta\gamma_{(m} \pd^{p}\mathbf{F}_{n)p})
\\
\delta_{\zeta} H_{mnp} &= 3\bx (\zeta\gamma_{[m}\mathbf{F}_{np]})
- 3  (\zeta\gamma_{[mn}\pd_{p]}\mathbf{i})   
+ \frac{1}{2} (\zeta\gamma_{[mn}\pd_{p]} \pd_{\ell} \mathbf{F}^{\ell})
\\
\delta_{\zeta} \mathbf{F}_{mn}^{\al} &=  
-2(\gamma^{rs}\zeta)^{\al} R_{mrsn}
+4 (\gamma^{rp}\zeta)^{\al}\pd_{[n} C_{m]rp} 
-\frac{2}{3} (\pd_{[n}\gamma_{m]rps}\zeta)^{\al}C^{rps}
\\
\delta_{\zeta} \mathbf{F}_{m\al} &=  
+ 2(\gamma^{n}\zeta)_{\al}R_{mn}
-2 (\gamma^{lnp}\zeta)_{\al}\pd_{l} C_{mnp} +
\frac{1}{3} (\gamma_{lmnps}\zeta)_{\al}\pd^{l} C^{nps}
\nonumber \\
&\quad +4 (\gamma^{n}\zeta)_{\al} \pd^p C_{mnp} - (\gamma_{mps}\zeta)_{\al}\pd_{n}C^{nps}
\\
\delta \slashed{\mathbf{F}}^{\bt} &= 2\zeta^{\bt} R
-6 (\gamma^{np}\zeta )^{\bt}\pd^{m}C_{npm}
\end{align}
\end{subequations}

\subsection{Supersymmetry for $(h_{mn}, t, \mathbf{i} ,\mathbf{d}) $}

Let us consider the system:

\begin{align} 
\mathbf{S} &= - \int d^{10} x \Bigg(
\frac{1}{2} h^{mn}\bx\left( R_{mn}
-\frac{1}{2}\eta_{mn}R\right) - tR
\nonumber \\
&\hspace{3.5cm}
+ \frac{1}{2}\mathbf{d}^{m\al}\bx\left( \mathbf{F}_{m\al}
-\frac{1}{2}(\gamma_{m})_{\al\bt}\slashed{\mathbf{F}}^{\bt}\right)
+\frac{1}{2}(\slashed{\mathbf{F}})^{\al}\mathbf{i}_{\al} \Bigg)
\end{align} 

\noindent
such that the

\paragraph{$S_{NS}$ variation} is given by

\begin{align*}
\delta_{\zeta} \left( -t R \right) &= - \zeta^{\al} \mathbf{i}_{\al}R
- 2 t\zeta^{\al}\pd_{p} \mathbf{F}_{\al}^{p}
\\
\\
\delta_{\zeta}\left[ \frac{1}{2} h^{mn}\bx\left( R_{mn}  -\frac{1}{2}\eta_{mn}R\right)  \right]
&= 2(\zeta \gamma^m  \mathbf{d}^{n})\bx\left( R_{mn}  -\frac{1}{2}\eta_{mn}R\right)
\end{align*}

\noindent
and the

\paragraph{$S_{R}$ variation} is given by

\begin{align*}
\delta_{\zeta} \left(\frac{1}{2}\mathbf{d}^{m\al}\bx\left( \mathbf{F}_{m\al}
-\frac{1}{2}(\gamma_{m})_{\al\bt}\slashed{\mathbf{F}}^{\bt}\right) \right)
&=  2\mathbf{d}^{m\al}\bx\left((\zeta\gamma^n)_{\al}R_{mn}-\frac{1}{2}(\zeta\gamma_{m})_{\al}R\right) \\
&=-2 (\zeta\gamma^n\mathbf{d}^{m})\bx\left(R_{mn}
-\frac{1}{2}\eta_{mn}R\right)
\end{align*}

\begin{align*}
\delta_{\zeta} \left( \frac{1}{2} (\slashed{\mathbf{F}})^{\al}\mathbf{i}_{\al}  \right)
&=  \zeta^{\al} R \mathbf{i}_{\al} +
(\slashed{\mathbf{F}})^{\al}\slashed{\pd}_{\al\bt}\zeta^{\bt}t\\
&=  \zeta^{\al} \mathbf{i}_{\al}R +
2\zeta^{\al} (\pd_{p} \mathbf{F}_{\al}^{p} ) t + \pd (...)
\end{align*}

\noindent
where we have used \eqref{susyfs} and $(\slashed{\pd}\slashed{\mathbf{F}}\zeta) = 2 (\zeta \pd_{p} \mathbf{F}^{p}) $.
It is clear that the sum of all terms cancels and invariance of this system is stablished. 

\subsection{Supersymmetry for $(H_{mnp},C_{mnp},\mathbf{d}_{m}^{\al},\mathbf{i}_{\al})$  }

It remains for consideration the following system:

\begin{align}
\mathbf{S} &= - \int d^{10} x \Bigg(
-C^{mnp}H_{mnp} + \frac{1}{2}C^{mnp}\left(\bx C_{mnp}  - \frac{1}{2}\pd_{[p}\pd^{r}C_{mn]r} \right) 
\nonumber \\
&\hspace{4.0cm} +
\frac{1}{2}\mathbf{d}^{m\al}\bx\left( \mathbf{F}_{m\al}
-\frac{1}{2}(\gamma_{m})_{\al\bt}\slashed{\mathbf{F}}^{\bt}\right)
+\frac{1}{2}(\slashed{\mathbf{F}})^{\al}\mathbf{i}_{\al} \Bigg)
\end{align}

\noindent
In order to check supersymmetric invariance we have to gather all independent combination
of gamma matrices $(\gamma^m,\gamma^{mn}$, $\gamma^{mnp},\gamma^{mnpp},\gamma^{mnpqr})$.
So consider the

\paragraph{$S_{NS}$ variation:}

\begin{align*}
\delta_{\zeta} (-C^{mnp}H_{mnp}) & = 
+3\left[  (\zeta\gamma_{tmn}\mathbf{F}^{t}_{\; p}) + 
(\zeta\gamma_{m}\mathbf{F}_{n p}) \right] H^{mnp} 
-3(\zeta \gamma_{m}\mathbf{F}_{np}) \bx C^{mnp} \\
&\quad-3  (\zeta \gamma_{mn}\mathbf{i}) \pd_{p}C^{mnp} 
-\frac{1}{2} (\zeta\gamma_{mn} \pd_{p} \mathbf{F}^{p})\pd_{p}C^{mnp} + \pd(\dots)
\end{align*}

\begin{align*}
\delta_{\zeta} \left[\frac{1}{2}C^{mnp}\left(\bx C_{mnp}  
-\frac{1}{2}\pd_{[p}\pd^{r}C_{mn]r} \right)\right] 
=& - 3 \left[  (\zeta\gamma_{tmn}\mathbf{F}^{t}_{\; p}) 
+ (\zeta\gamma_{m}\mathbf{F}_{n p})\right]\bx C^{mnp}\\
&+ \frac{1}{2} \left[ (\zeta\gamma_{mn}\pd^p\mathbf{F}_{p})+ 
(\zeta\gamma_{m}\pd_n\slashed{\mathbf{F}})  \right]\pd_{r}C^{mnr} 
+ \pd(\dots)
\end{align*}

\noindent
and the

\paragraph{$S_{R}$ variation:}

\begin{align*}
\delta_{\zeta}\left(\frac{1}{2}\slashed{\mathbf{F}}\mathbf{i}\right) &= 
-3(\gamma^{np}\zeta )^{\bt}\pd^{m}C_{npm} \mathbf{i}_{\bt}  
-\frac{1}{2}\slashed{\mathbf{F}}^{\al}(\gamma^{mnp}\zeta)_{\al}H_{mnp}  
+\frac{1}{6}\slashed{\mathbf{F}}^{\al}  (\gamma^{mnp}\zeta)_{\al} \bx C_{mnp}\\
&=+3(\zeta\gamma^{nm}\mathbf{i} )\pd^p C_{nmp} + \\
&\quad-\left[\frac{1}{6}  (\zeta\gamma^{mnpts}\mathbf{F}_{ts})\bx C_{mnp} 
-  (\zeta\gamma^{tmn}\mathbf{F}^{\;\,p}_{t}) \bx C_{mnp}
- (\zeta\gamma^{m}\mathbf{F}^{np}) \bx C_{mnp}\right]\\
&\quad+\left[\frac{1}{2}  (\zeta\gamma^{mnpts}\mathbf{F}_{ts})H_{mnp} 
- 3 (\zeta\gamma^{tmn}\mathbf{F}^{\;\,p}_{t}) H_{mnp}
-3 (\zeta\gamma^{m}\mathbf{F}^{np}) H_{mnp}\right]
\end{align*}

\begin{align*}
&\delta_{\zeta}\left(\frac{1}{2}\mathbf{d}^{\al}_{m}\bx  \left( \mathbf{F}_{\al}^{m}
-\frac{1}{2}(\gamma^{m})_{\al\bt}\slashed{\mathbf{F}}^{\bt}\right)\right) =\\
&\qquad\qquad=
\left(-2(\gamma^{np}\zeta)^{\al} C_{mnp} +
\frac{1}{3}(\gamma_{mnps}\zeta)^{\al} C^{nps}\right)\bx\left( \mathbf{F}^{m}_{\al}
-\frac{1}{2}(\gamma^{m})_{\al\bt}\slashed{\mathbf{F}}^{\bt}\right)\\
&\qquad\qquad= +2 (\mathbf{F}^{\;\;m}_{l} \gamma^{lnp}\zeta)\bx C_{mnp} 
-4(\mathbf{F}^{mn}\gamma^{p}\zeta)\bx C_{mnp} +\\
&\qquad\qquad -\frac{1}{3} (\mathbf{F}_{lm}\gamma^{lmnps}\zeta)\bx C_{nps} 
- (\mathbf{F}_{m}^{\;\;n}\gamma^{psm}\zeta)\bx C_{nps}+\\
&\qquad\qquad-\bigg[\frac{1}{6}  (\zeta\gamma^{mnpts}\mathbf{F}_{ts})\bx C_{mnp} 
-  (\zeta\gamma^{tmn}\mathbf{F}^{\;\,p}_{t}) 
\bx C_{mnp} -  (\zeta\gamma^{m}\mathbf{F}^{np}) \bx C_{mnp}\bigg]
\end{align*}

\noindent
Recall that the $\gamma^{mnpqr}$ is symmetric and $\gamma^{mnp}$ is 
antisymmetric under the spinor indices. Gathering all independent terms 
we confirm the system is supersymmetric.

\vspace{.5cm}
\noindent
\paragraph{Acknowledgements.}
We would like to thank Oliver Schlotterer and Nathan Berkovits for useful discussions,
Andrei Mikhailov and Nathan Berkovits for helpful comments on the manuscript.
ML thanks FAPESP grant 2016/16824-0 for financial support
and HF would like to thank CAPES grant 33015015001P7 for financial support.

\appendix
\numberwithin{equation}{section}

\section{Ramond sector, cocycles and Gamma matrices.} \label{appA}

Spinor indices in $10$ dimensions can be distinguished
between chiral and anti-chiral. We denote chiral indices by
undotted greek letters, $\alpha$,
while anti-chiral indices are represented by dotted greek letters,
$\da$. Both run from $1$ to $16$.
Spinor indices are $5$-dimensional vector representations of $u(5)$:

\begin{equation}
\dot{\alpha} = \frac{1}{2}
\begin{pmatrix}
- & - & - & - & - \\
- & - & - & + & + \\
- & + & + & + & +
\end{pmatrix}
\quad \text{and} \quad
\beta = \frac{1}{2}
\begin{pmatrix}
+ & + & + & + & + \\
+ & + & + & - & - \\
+ & - & - & - & -
\end{pmatrix}.
\end{equation}

\noindent
where an anti-chiral index, $\dot{\alpha}$, must have an even number of plus signs,
and a chiral index, $\beta$, must have an odd number of plus signs.
Each of these combinations has 16 independent components represented as
$\bf{16} = \bf{1} + \bf{10} + \bf{5}$.

\subsection{The Ramond Sector.}

The Ramond sector of the Ambitwistor string is defined
by the antiperiodic boundary conditions of $\psi^{m}$:

\begin{equation}
\psi^{m} ( e^{2 \pi i} z ) = - \psi^{m} (z).
\end{equation}

\noindent
We follow\cite{Friedan:1985ge} and
implement these boundary conditions via spin fields.
That is, we have a conformal primary $S(z)$
that twists a periodic $\psi$:

\begin{equation}
\psi^{m} \left( z + (w - z) e^{2 \pi i} \right) S(z) =
- \psi^{m} \left( w \right) S(z).
\end{equation}

\noindent
This implies that a state $ | \alpha \rangle$
created from the vacuum $ | 0 \rangle$ via

\begin{equation}
| \alpha \rangle = S^{\alpha} (0) | 0 \rangle
\end{equation}

\noindent
should transform as a spacetime spinor. Notice
that, due to the presence of $S$ forcing $\psi$ to be in
the Ramond sector, this state must
belong to an irreducible representation of the zero-mode
Clifford algebra of $\psi^{m}$:
$\{ \psi^{m}_{0} , \psi^{n}_{0} \} = \eta^{mn}$,
which implies

\begin{equation} \label{bla}
\psi^{m}_{0} | \alpha \rangle =
\frac{1}{\sqrt{2}} \Gamma^{m \, \alpha}_{\bd} | \bd \rangle.
\end{equation}

\subsection{Bosonization and cocycles.}

Because $S^{\alpha}$ twists the boundary conditions of $\psi^{m}$,
the system is not free and OPE's are difficult to compute.
Bosonization is a technique that allows us to deal with free fields only.
Bosonization assigns for a pair of complex fermions one chiral boson, which
means that we have to break manifest $so(10)$ invariance down to $u(5)$.

\paragraph{Spin Fields.}
The bosonization of spin fields is given by

\begin{equation} \label{bosos}
S^{\alpha}(z) = \exp \bigg( \alpha \cdot \phi (z) \bigg) c_{\alpha}
\end{equation}

\noindent
where $\alpha$ is a chiral spinor index.  
The same expression is valid for anti-chiral spin fields
by just replacing $\alpha$ for $\ad$.
The factor $c_{\alpha}$ is a cocycle phase that guarantees
the correct anticommutation relations. 

\paragraph{Cocycles.}
The anticommuting fermionic algebra is reproduced in the bosonic system via the 
\href{https://en.wikipedia.org/wiki/Baker–Campbell–Hausdorff_formula}{Baker-Campbell-Hausdorff formula}:

\begin{equation}
e^{\phi(z)} e^{\pm \phi(z')} = e^{\pm \phi(z')} e^{\mp \phi(z')} e^{\phi(z)} e^{\pm \phi(z')} = - e^{\pm \phi(z')} e^{\phi(z)}
\end{equation}

\noindent provided for $|z'| = |z|$ we have

\begin{equation}
\bigg[ \phi(z'), \phi(z) \bigg]  = \pm i \pi
\quad \texttt{ which implies } \quad
\phi(z) \phi (0) \sim \ln z
\end{equation}

\noindent
Now, if we are given more than one pair of fermions, they
won't naturally anticommute because $\left[ \phi_{i} , \phi_{j} \right] = 0$.
This is corrected by the introduction of cocycles\cite{Kostelecky:1986xg}:

\begin{itemize}
\item Order all bosons of the theory: $\phi_{i}$ where $i=1,\ldots,N$;

\item Then multiply each exponential by a factor $(-)^{N_{1} + \ldots + N_{i-1}}$,
where $N_{i}$ is the fermion number operator:

\begin{equation}
N_{i} = - \oint \frac{\dd z}{2 \pi i } \, \overline{\psi}_{i} \psi_{i} = \oint \frac{\dd z}{2 \pi i} \, \partial \phi_{i}.
\end{equation}
\end{itemize}

\noindent
For example, if we consider two pairs of fermions, the bosonization becomes

\begin{equation}
\psi_{1} = e^{\phi_{1}}, \quad \overline{\psi}_{1} = e^{-\phi_{1}}
\end{equation}

\noindent
with

\begin{equation}
\psi_{2} = e^{\phi_{2}} (-)^{N_{1}}, \quad \overline{\psi}_{2} = e^{-\phi_{2}} (-)^{N_{1}}
\end{equation}

\noindent
where now $\psi_{1}$ and $\psi_{2}$ anticommute 

\begin{equation}
e^{\phi_{1}} e^{\phi_{2}} (-)^{N_{1}} = e^{\phi_{2}} e^{\phi_{1}} (-)^{N_{1}} = e^{\phi_{2}} (-)^{N_{1}} (-)^{-N_{1}} e^{\phi_{1}} (-)^{N_{1}} = - e^{\phi_{2}} (-)^{N_{1}} e^{\phi_{1}}
\end{equation}

\noindent
provided

\begin{equation}
\left[ N_{i} , e^{n \phi_{j}} \right] = n \delta_{ij} e^{n \phi_{j}}.
\end{equation}

\noindent
Thus, for more than one pair of fermions, we need to introduce the cocycle phase factors:

\begin{equation}
c_{i} = (-)^{N_{1} \, + \, \ldots \, + \, N_{i-1}}.
\label{factor}
\end{equation}

Consider the vector

\begin{equation}
\partial \phi = \left( N_{1}, N_{2}, \ldots, N_{5} \right)
\end{equation}

\noindent
then the cocycle factor can be written as

\begin{equation}
c_{\pm e_{i}} = \exp \left[ \pm i \pi \langle e_{i} M \partial \phi \rangle \right]
\end{equation}

\noindent
where $e_{i}$ is $1$ in the $i$th component and zero elsewhere, $\langle \, \, \, \rangle$ is a matrix inner product
and $M$ is a lower triangular matrix with entries $\pm 1$:

\[ M =
\begin{pmatrix}
  0 & 0 & 0 & 0 & 0 & 0 \\
  1 & 0 & 0 & 0 & 0 & 0 \\
  1 & 1 & 0 & 0 & 0 & 0 \\
  -1 & 1 & -1 & 0 & 0 & 0 \\
  1 & 1 & 1 & 1 & 0 & 0 \\
  -1 & -1 & -1 & -1 & -1 & 0 \\
\end{pmatrix}.
\]

\noindent
The signs of $M$ are arbitrary at this point,
but they can be specified studying the charge conjugation matrix\cite{Kostelecky:1986xg}.

The cocycle factors of spin fields,
$c_{\al}$ and $c_{\ad}$, are given the following expressions:

\begin{equation}
c_{\al} = \exp \left[ i \pi \langle \al M \partial \phi \rangle \right]
\quad \texttt{and} \quad
c_{\ad} = \exp \left[ i \pi \langle \ad M \partial \phi \rangle \right]
\end{equation}

\paragraph{Gamma Matrices.}
To motivate the construction of gamma matrices and show how cocycles work, let us consider the OPE between
$\psi^{i}$ and $S^{\al}$. Using expressions \eqref{pbosonization} and \eqref{bosos}
we have to compute the OPE of $ e^{\phi_{i} (z) } c_{i} $ with $e^{\alpha \phi(\mw) } c_{\alpha} $.
Notice that $c_{i}$ will pass through $e^{\alpha \phi}$ and due to Baker-Campbell-Hausdorff we obtain
an extra phase:

\begin{equation}
c_{i} e^{\alpha \phi} = e^{i \pi \langle e_{i} M \partial \phi \rangle} e^{\alpha \phi} =
e^{i \pi \langle e_{i} M \alpha \rangle} e^{\alpha \phi} c_{i}
\end{equation}

\noindent
so that our OPE becomes

\begin{equation}
e^{\phi_{i} (z) } c_{i} \,\, e^{\alpha \phi(\mw) } c_{\alpha}
\sim
(z - \mw)^{\alpha \cdot e_{i}} \, e^{i \pi \langle e_{i} M \al \rangle} e^{(e_{i} + \al) \phi} c_{i + \alpha}.
\end{equation}

\noindent
Notice that we obtain a branch-cut if $\al \cdot e_{i} = \al_{i} = -1/2$
which in turn implies that the sum $e_{i} + \al$ must be an anti-chiral index $\bd$.
Therefore given

\begin{equation}
e^{\phi_{i} (z) } c_{i} \,\, e^{\alpha \phi(\mw) } c_{\alpha}
\sim
(z - \mw)^{-1/2} \, e^{i \pi \langle e_{i} M \al \rangle} e^{\bd \phi} c_{\bd},
\end{equation}

\noindent
we see that it becomes natural to define the
gamma matrices as

\begin{subequations}
\begin{equation}
(\Gamma^{j})^{\, \beta}_{\ad} = \sqrt{2} \delta \left( e_{j} + \beta - \ad \right) e^{i \pi \langle e_{j} M \ad \rangle}
\end{equation}

\noindent
and

\begin{equation}
(\Gamma^{j} )^{\, \bd}_{\alpha} = \sqrt{2} \delta \left( e_{j} + \bd - \alpha \right) e^{i \pi \langle e_{j} M \alpha \rangle}
\end{equation}
\end{subequations}

\noindent
giving us the final result:

\begin{equation}
\psi^{i} (z) S^{\alpha}(\mw) \sim
\frac{1}{\sqrt{2}}
\frac{ \Gamma^{i \, \alpha}_{\bd} {S}^{\bd} (\mw) }{(z - \mw)^{1/2}}.
\end{equation}

The explicit representation is written in terms of the Pauli-matrices via

\begin{equation}
\Gamma^{\pm e_{j}} = \left( \pm i \right)^{j-1} \sqrt{2} \left( \sigma^{3} \otimes \right)^{j-1} \sigma^{\mp} \left(\otimes 1 \right)^{5-j}
\end{equation}

\noindent
and one can convert between $u(5)$ and covariant $so(10)$ using

\begin{subequations}
\begin{equation}
\Gamma^{2j - 1} = \frac{1}{\sqrt{2}} \left( \Gamma^{e_{j}} + \Gamma^{-e_{j}} \right)
\end{equation}

\noindent
and

\begin{equation}
\Gamma^{2j} = \frac{i}{\sqrt{2}} \left( \Gamma^{e_{j}} - \Gamma^{-e_{j}} \right)
\end{equation}
\end{subequations}

\noindent
Notice that in our construction, the notation $\gamma^{\mu}$ is reserved for the
symmetric gamma matrices:

\begin{subequations}
\begin{align}
&\gamma^{\mu}_{\alpha \beta} = \Gamma^{\mu \, \bd}_{\alpha} C_{\bd \beta}
\\
\nonumber \\
&\gamma^{\mu \, \alpha \beta} = \Gamma^{\mu \, \alpha}_{\bd} C^{\bd \beta}
\end{align}
\end{subequations}

\noindent
as it is common in the literature.
In above equations, $C$ denotes the charge conjugation matrix
which is the next topic in our discussion.

\paragraph{Charge Conjugation Matrix.}
We define $C$ as

\begin{subequations}
\begin{equation}
C^{\beta \bd} = \delta \left( \beta + \bd \right) e^{i \pi \beta M \bd}
\end{equation}

\noindent
and

\begin{equation}
C^{\bd \beta} = - \delta \left( \bd + \beta \right) e^{i \pi \bd M \beta}
\end{equation}
\end{subequations}

\noindent
and with these convetions we have $C^{\beta \bd} = C^{\bd \beta}$.
These expressions can be motivated by studying the OPE of $S^{\al}$ and $S^{\bd}$.

It is also common to use only undotted indices when describing spinors in 10d.
Charge matrices act as metrics on the spinor space and can remove all dotted indices.
For us all spinors are defined with upper indices and then
anti-chiral ones are written as

\begin{equation}
{S}_{\beta} = C_{\beta \bd} {S}^{\bd}.
\end{equation}

\noindent
This notation is used together with the symmetric gamma representation.


\begin{thebibliography}{10}

\bibitem{Mason:2013sva}
L.~Mason and D.~Skinner,
``Ambitwistor strings and the scattering equations,''
JHEP {\bf 1407}, 048 (2014)
\href{https://arxiv.org/abs/1311.2564}{[arXiv:1311.2564 [hep-th]]}.

\bibitem{Cachazo:2014}
F.~Cachazo, S.~He and E.~Y.~Yuan,
``Scattering of Massless Particles in Arbitrary Dimensions,''
Phys.\ Rev.\ Lett.\  {\bf 113}, no. 17, 171601 (2014).
\href{https://arxiv.org/abs/1307.2199}{[arXiv:1307.2199 [hep-th]]}.


\bibitem{Berkovits:2018jvm}
N.~Berkovits and M.~Lize,
``Field theory actions for ambitwistor string and superstring,''
JHEP {\bf 1809}, 097 (2018)
doi:10.1007/JHEP09(2018)097
[arXiv:1807.07661 [hep-th]].

\bibitem{Fang:1978wz}
J.~Fang and C.~Fronsdal,
``Massless Fields with Half Integral Spin,''
Phys.\ Rev.\ D {\bf 18}, 3630 (1978).
doi:10.1103/PhysRevD.18.3630

\bibitem{Friedan:1985ge}
D.~Friedan, E.~J.~Martinec and S.~H.~Shenker,
``Conformal Invariance, Supersymmetry and String Theory,''
Nucl.\ Phys.\ B {\bf 271}, 93 (1986).
doi:10.1016/0550-3213(86)90356-1, 10.1016/S0550-3213(86)80006-2

\bibitem{Kostelecky:1986xg}
V.~A.~Kostelecky, O.~Lechtenfeld, W.~Lerche, S.~Samuel and S.~Watamura,
``Conformal Techniques, Bosonization and Tree Level String Amplitudes,''
Nucl.\ Phys.\ B {\bf 288}, 173 (1987).
doi:10.1016/0550-3213(87)90213-6

\bibitem{Koh:1987hm}
I.~G.~Koh, W.~Troost and A.~Van Proeyen,
``Covariant Higher Spin Vertex Operators in the Ramond Sector,''
Nucl.\ Phys.\ B {\bf 292}, 201 (1987).
doi:10.1016/0550-3213(87)90642-0


\end{thebibliography}
\end{document}